\documentstyle[12pt,epsfig]{article}
  \newcommand{\ccaption}[2]{
    \begin{center}
    \parbox{0.85\textwidth}{
      \caption[#1]{\small{{#2}}}
      }
    \end{center}
    }
\def\be{\begin{equation}}
\def\ee{\end{equation}}
\def\bc{\begin{center}}
\def\ec{\end{center}}
\def\bea{\begin{eqnarray}}
\def\eea{\end{eqnarray}}

\def\nn{\nonumber}

\def\co{{\cal O}}

\def\ctg{\cos \theta_{\gamma}}
\def\ctgl{\cos \theta_{g}}
\def\ctq{\cos \theta_{q}}
\def\dab{{\delta^{\alpha}_{\beta}}}
     
\def\dmd{\partial_\mu}

\def\dslash{\not{\! \partial}}
\def\Dslash{\not{\!\! D}}

\def\et{E_T}

\def\etmin{E_T^{min}}

\def\ev{{\rm \; eV}}   
\def\fb{{\rm \; fb}}
\def\frac#1#2{{#1 \over #2}}
\def\gev{{\rm \; GeV}}
\def\glu{{(A_{\mu})^{\alpha}_{\beta}}}

\def\grav{\tilde{G}}
\def\gsim{\raisebox{-3pt}{$\>\stackrel{>}{\scriptstyle\sim}\>$}}

\def\met{\not{\!\! E}_T}

\def\mpl{M_{\rm P}}
\def\ov{\overline}
\def\pb{{\rm \; pb}}

\def\psis{\tilde{G}}
\def\psisb{\overline{\tilde{G}}}

\def\qal{{q^{\alpha}}}
\def\qbal{{\overline{q}_{\alpha}}}
\def\qbet{{q^{\beta}}}

\def\simlt{\stackrel{<}{{}_\sim}}
\def\simgt{\stackrel{>}{{}_\sim}}
\def\sqtg{\sin^2 \theta_{\gamma}}
\def\sqtgl{\sin^2 \theta_{g}}

\def\tev{{\rm \; TeV}}     
\def\tevI{TeV~I}     
\def\tevII{TeV~II}

\def\xg{x_{\gamma}}
\def\xgl{x_{g}}
\def\xq{x_{q}}

\catcode`@=11
\def\marginnote#1{}
\newcount\hour
\newcount\minute
\newtoks\amorpm
\hour=\time\divide\hour by60
\minute=\time{\multiply\hour by60 \global\advance\minute by-\hour}
\edef\standardtime{{\ifnum\hour<12 \global\amorpm={am}%
        \else\global\amorpm={pm}\advance\hour by-12 \fi
        \ifnum\hour=0 \hour=12 \fi
        \number\hour:\ifnum\minute<10 0\fi\number\minute\the\amorpm}}
\edef\militarytime{\number\hour:\ifnum\minute<10 0\fi\number\minute}
\def\draftlabel#1{{\@bsphack\if@filesw {\let\thepage\relax
   \xdef\@gtempa{\write\@auxout{\string
      \newlabel{#1}{{\@currentlabel}{\thepage}}}}}\@gtempa
   \if@nobreak \ifvmode\nobreak\fi\fi\fi\@esphack}
        \gdef\@eqnlabel{#1}}
\def\@eqnlabel{}
\def\@vacuum{}
\def\draftmarginnote#1{\marginpar{\raggedright\scriptsize\tt#1}}
\def\draft{\oddsidemargin 0.0truein
        \def\@oddfoot{\sl preliminary draft \hfil
        \rm\thepage\hfil\sl\today\quad\militarytime}
        \let\@evenfoot\@oddfoot \overfullrule 3pt
        \let\label=\draftlabel
        \let\marginnote=\draftmarginnote
   \def\@eqnnum{(\theequation)\rlap{\kern\marginparsep\tt\@eqnlabel}%
\global\let\@eqnlabel\@vacuum}  }
\catcode`@=12
\begin{document}
\begin{titlepage}
\vspace*{-1cm}
\phantom{bla}
\hfill{CERN-TH/98-11}
\\
\phantom{bla}
\hfill{DFPD~98/TH/04}
\\
\vskip 1.0cm
\bc
{\Large \bf Signals of a superlight gravitino at hadron
colliders \\
when the other superparticles are heavy \footnote{Work
supported in part by the European Commission TMR
Programme ERBFMRX-CT96-0045.}}
\ec
\vskip 0.8cm
\bc
{\large
Andrea Brignole $^{a,b,}$\footnote{e-mail address: 
brignole@pd.infn.it},
Ferruccio Feruglio $^{a,b,c,}$\footnote{e-mail address: 
feruglio@pd.infn.it}, }
\ec         
\bc
{\large
  Michelangelo L. Mangano $^{c,}$\footnote{On leave of absence from:
  Istituto Nazionale di Fisica Nucleare, Sezione di Pisa,
  I-56100  Pisa, Italy.}$^{,}$\footnote{e-mail address: mlm@vxcern.cern.ch}
  and
Fabio Zwirner $^{a,b,}$\footnote{e-mail address: 
zwirner@pd.infn.it}
}
\ec
\vskip 0.5cm
\bc
$^a$ 
Istituto Nazionale di Fisica Nucleare, Sezione di Padova, 
I-35131 Padua, Italy
\\
$^b$
Dipartimento di Fisica, Universit\`a di Padova,
I-35131 Padua, Italy
\\
$^c$
Theory Division, CERN, CH-1211 Geneva 23, Switzerland
\ec
\vskip 0.5cm
\begin{abstract}
\noindent
If the gravitino $\grav$ is very light and all the other
supersymmetric particles are above threshold, supersymmetry
may still be found at colliders, by looking at processes with
only gravitinos and ordinary particles in the final state.
We compute here the cross-sections for some distinctive signals
at hadron colliders: photon plus missing energy, induced by  
$q \ov{q} \to \grav \grav \gamma$, and jet plus missing energy, 
induced by $q \ov{q} \to \grav \grav g$, $q g \to \grav \grav q$,
$\ov{q} g \to \grav \grav \ov{q}$ and $g g \to \grav \grav g$.
From the present Tevatron data, we estimate the bound $m_{3/2} > 
2.3 \times 10^{-5} \ev$ on the gravitino mass, corresponding to 
the bound $\sqrt{F} > 310 \gev$ on the supersymmetry-breaking scale. 
We foresee that the upgraded Tevatron and the LHC will be sensitive 
to values of $m_{3/2}$ up to $4.0 \times 10^{-5} \ev$ and $6.2 
\times 10^{-4} \ev$, corresponding to $\sqrt{F}$ up to $410 \gev$ 
and $1.6 \tev$, respectively. 
\end{abstract}
\begin{center}
{\em This revised version supersedes that published in Nucl. Phys. 
B526~(1998)~136, and contains important changes. The correction 
of a sign error modifies the relevant partonic cross-sections. 
The sensitivity to $\sqrt{F}$ ($m_{3/2}$) is only slightly weakened.}
\end{center}
\vfill{
CERN-TH/98-11
\newline
\noindent                              
January 1998}
\end{titlepage}
\setcounter{footnote}{0}
\vskip2truecm
\section{Introduction}
While there are strong theoretical motivations for supersymmetry, 
with supersymmetry-breaking mass splittings of the order of the 
electroweak scale (for reviews and references, see e.g. \cite{susy}),
there is at present no compelling argument to select a definite 
value of the supersymmetry-breaking scale $\sqrt{F}$ or, 
equivalently\footnote{We recall that, in a flat space-time, 
$F = \sqrt{3} \, m_{3/2} \mpl$, where $\mpl = (8 \pi G_N)^{-1/2} 
\simeq 2.4 \times 10^{18} \gev$ is the Planck mass. Note that
we take $F$ real and positive, which is not restrictive for the
present paper.}, the gravitino mass $m_{3/2}$.  
As reviewed in \cite{bfz3}, different possibilities should then 
be kept in mind when performing phenomenological analyses: 
they are characterized by a `heavy', `light' or `superlight' 
gravitino. In the heavy gravitino case, reactions involving the 
gravitino are never important for collider physics. In the light 
gravitino case, the gravitino can be relevant in the decays of other 
supersymmetric particles, if there is sufficient energy 
to produce the latter. In the superlight gravitino
case, also the direct production of gravitinos (with
or without other supersymmetric particles) can become
relevant. In this paper, we concentrate on the superlight 
gravitino case, where $\sqrt{F}$ can be close to the electroweak 
scale and, correspondingly, $m_{3/2}$ can be several orders of 
magnitude below the $\ev$ scale.

Many aspects of the superlight gravitino phenomenology at
colliders have been discussed long ago \cite{fayet}, 
and also more recently \cite{slgcol}. In all these papers, 
however, it was assumed that some other supersymmetric particle, 
for example a neutralino or one of the spin-0 partners of the 
gravitino, is light enough to be produced on-shell in some 
reaction. Here we take an orthogonal point of view: there 
may be experiments where the available energy is still
insufficient for the on-shell production of other
supersymmetric particles, but nevertheless sufficient to
give rise to final states with only gravitinos and ordinary
particles, at measurable rates. In a recent paper \cite{bfz4},
some of us considered the process $e^+ e^- \to \grav \grav 
\gamma$, which may give rise to a distinctive {\em photon + missing 
energy}  signal at $e^+ e^-$ colliders. In the present work, 
we study the possible signals of a superlight gravitino at hadron 
colliders, such as the Tevatron or the LHC. At the partonic level, 
we consider the following subprocesses:
\bea
q + \ov{q} & \longrightarrow & \grav + \grav + \gamma \, ,
\label{qqgam}
\\
q + \ov{q} & \longrightarrow & \grav + \grav + g \, ,
\label{qqglu}
\\
q + g & \longrightarrow & \grav + \grav + q \, ,
\label{qglu}
\\
\ov{q} + g  & \longrightarrow & \grav + \grav + \ov{q} \, ,
\label{qbarglu}
\\
g + g & \longrightarrow & \grav + \grav + g \, .
\label{gluglu}
\eea
The process of eq.~(\ref{qqgam}) corresponds to a 
{\em photon + missing transverse energy} ($\met$) signal, those of 
eqs.~(\ref{qqglu})--(\ref{gluglu}) to a 
{\em jet} + $\met$ signal. We compute the 
cross-section and the relevant angular distributions 
for the processes of eqs.~(\ref{qqgam})--(\ref{gluglu}), 
in the limit in which the supersymmetric particles 
of the Minimal Supersymmetric Standard Model (MSSM), 
such as squarks and gauginos, and all other 
exotic particles, such as the spin-0 partners of the
goldstino, are heavy. We then analyse the resulting 
phenomenology at the Tevatron ($p \ov{p}$, $\sqrt{S} =
1.8 \tev$, $L \sim 100 \pb^{-1}$), the upgraded 
Tevatron ($p \ov{p}$, $\sqrt{S} = 2 \tev$, $L \sim 
2 \fb^{-1}$), and the LHC ($pp$, $\sqrt{S} = 14 \tev$, 
$L \sim 10 \fb^{-1}$).                                  
We show how the present Tevatron data, in the absence of a 
signal over the Standard Model background, should allow us to 
establish the lower bound $\sqrt{F} \simgt 310 \gev$, or, 
equivalently, $m_{3/2} \simgt 2.3 \times 10^{-5} \ev$. 
These bounds are considerably better than the one estimated 
\cite{bfz4} from the present LEP data, $\sqrt{F} \simgt 200
\gev$ or $m_{3/2} \simgt 10^{-5} \ev$. We also estimate the 
future sensitivity of the upgraded Tevatron and the LHC: 
$\sqrt{F}$ up to $410 \gev$ and $1.6 \tev$, corresponding to
$m_{3/2}$ up to $4.0 \times 10^{-5} \ev$ and $6.2 \times
10^{-4} \ev$, respectively. In contrast with other collider bounds 
discussed in the literature \cite{slgcol}, those discussed 
in the present paper cannot be evaded by modifying the mass 
spectrum of the other supersymmetric particles: making some 
additional supersymmetric particle light leads in general 
to stronger bounds. Therefore, making use of our results,
an {\em absolute} lower bound on the gravitino mass can be 
established.

The paper is organized as follows. In sect.~2, we spell out 
and discuss the low-energy effective Lagrangian used for our 
calculations. In sect.~3, we compute the relevant differential 
cross-sections at the partonic level. In sects.~4 and 5, we 
discuss the phenomenology of the jet + $\met$ and 
$\gamma+\met$ signals at hadron colliders, 
emphasizing the implications of the present Tevatron data and 
the prospects for the upgraded Tevatron and the LHC. Finally, 
in sect.~6 we summarize and discuss our results.
                                      
\section{The effective Lagrangian}
For the theoretically oriented readers, we summarize the
framework of our calculation. We are considering processes
whose typical energies are much larger than the gravitino 
mass, but smaller than the masses of all the other particles
not belonging to the Standard Model.  We can then work
with the low-energy effective theory defined by the
following two-step procedure. We start from a
generic supergravity Lagrangian, assuming that supersymmetry 
is spontaneously broken with a light gravitino, and we take 
the appropriate low-energy limit: $M_P \to \infty$ with $\sqrt{F}$ 
fixed. In accordance with the supersymmetric equivalence 
theorem \cite{equiv}, gravitational interactions are 
consistently neglected in this limit, and we end up with an 
effective (non-renormalizable) theory with linearly realized, 
although spontaneously broken, global supersymmetry, whose 
building blocks are the chiral and vector supermultiplets
containing the light degrees of freedom, including the goldstino. 
To simplify the discussion, we assume that such a theory has
pure $F$-breaking and negligible higher-derivative terms.
Since we are interested in the case where the available energy 
is smaller than the supersymmetry-breaking mass splittings,
we perform a second step and move to a `more effective' theory, 
by explicitly integrating out the heavy superpartners in 
the low-energy limit. The only degrees of freedom left are then 
the goldstino and the Standard Model particles, and supersymmetry 
is non-linearly realized \cite{nl,bfzeegg,clnew}. 

The derivation of the low-energy effective Lagrangian, to be used
as the starting point for our calculations, proceeds exactly as in
\cite{bfz4}. The only differences are that now the unbroken gauge 
group is $SU(3)_C \times U(1)_{em}$ and that we are dealing with 
quarks instead of leptons. In analogy with \cite{bfz4}, we neglect
the light quark masses and any possible mixing in the squark sector.
Denoting with $q^{\alpha}$ the light quark fields (where $\alpha=
1,2,3$ and $q=u,d,s,c,b$), and with $\glu \equiv A_\mu^A 
(T^A)^\alpha_\beta$ the gluon fields (where $A=1,\ldots,8$ 
and $T^A$ are the $SU(3)_C$ generators), the covariant derivative 
with respect to $SU(3)_C \times U(1)_{em}$ reads\footnote{Notice 
the sign change with respect to the first version of this paper,
following the analogous sign change in ref.~\cite{bfz4}.}:
\be
\label{cov}
\left( D_{\mu} \right)^\alpha_\beta = 
\left( \dmd + i e Q_q A_{\mu} \right) 
\delta^\alpha_\beta + i g_S \glu  \, . 
\ee
Focusing only on the terms relevant for our calculations, we obtain:
\be 
{{\cal L}}_{eff} =
- {1\over 4} F_{\mu\nu} F^{\mu\nu}
- {1\over 4} F_{\mu\nu}^A F^{\mu\nu \, A}
+ i \sum_q \qbal (\Dslash)^\alpha_\beta \qbet 
+ {i\over 2} \psisb \dslash \psis
+ \sum_{i=1}^4 {\cal O}_i \, .
\label{leff}
\ee
The terms $\co_i$ $(i=1,\dots,4)$ are local operators, bilinear in the
goldstino and generated by the exchange of massive particles in the 
large-mass limit. Their field-dependent parts have mass dimension 
$d=8$, and the corresponding coefficients scale as $1/F^2$. The
operator $\co_1$ involves only vector bosons (photons or gluons) 
and goldstinos, and is generated \cite{bfz3} by integrating over 
the spin-0 partners of the goldstino and the gauginos:
\be
{\co_1} = 
- {i \over 64 F^2} 
\left(
\psisb \left[ \gamma^\mu , \gamma^\nu \right]
F_{\mu \nu} \dslash \left[ \gamma^\rho , \gamma^\sigma 
\right] F_{\rho \sigma} \, \psis
+
\psisb \left[ \gamma^\mu , \gamma^\nu \right]
F_{\mu \nu}^A (\Dslash)^{AB} \left[ \gamma^\rho , 
\gamma^\sigma \right] F_{\rho \sigma}^B \, \psis
\right)
\, ,
\label{elleuno}
\ee
where $\Dslash$ is constructed with the covariant derivative
in the adjoint representation. The operator $\co_2$ is a 
four-fermion interaction among quarks and goldstinos, recovered 
by combining a contact term in the original Lagrangian with 
contributions originating from squark exchanges~\cite{bfzeegg}:
\be
{\co_2} = - {1 \over 2 F^2} \left\{ \qbal
\psis \Box \left( \psisb \qal \right) -
\qbal \gamma_5 \psis \Box \left( \psisb
\gamma_5 \qal \right) \right\} \, .
\label{elledue}
\ee
\addtocounter{footnote}{-1}
The operator $\co_3$ is generated by attaching a photon or a
gluon to a squark exchanged among quarks and 
goldstinos\footnotemark:
\be
{\co_3} = - {i \over F^2} \sum_q 
\left[ e Q_q A^{\mu} \dab + g_s \glu \right]
\left\{ \qbal \psis \partial_\mu \left( \psisb \qbet
\right) - \qbal \gamma_5 \psis \partial_\mu
\left( \psisb \gamma_5 \qbet \right) \right\} \, .
\label{elletre}
\ee
The operators $\co_2$ and $\co_3$ are not gauge
invariant, but their sum is contained in a gauge-invariant
combination, as can easily be verified. Finally, the operator 
$\co_4$ is a gauge-invariant contact term that directly 
contributes (as $\co_3$) to the scattering amplitudes of
eqs.~(\ref{qqgam})--(\ref{qbarglu}).
It originates from a combined squark and gaugino 
exchange:
\bea
{\co_4} & = & - {i \over 8 F^2} 
\left[ e Q_q F_{\mu \nu} \dab 
+ g_S F_{\mu \nu}^A (T^A)^\alpha_\beta \right]
\left\{ \psisb \left[ \gamma^\mu , \gamma^\nu \right]
\left( \qbet \qbal \psis - \gamma_5
\qbet \qbal \gamma_5 \psis \right) \right.
\nonumber \\ 
& & \phantom{i e Q_e \over 8 F^2}
\left. + \left( \psisb \qbet \qbal - \psisb
\gamma_5 \qbet \qbal \gamma_5 \right)
\left[ \gamma^\mu , \gamma^\nu \right]
\psis \right\} \, .
\label{ellequa}
\eea   

As a final remark, we should mention that the processes 
considered in the present paper were also considered in 
\cite{nach}, in the same kinematical limit, but relying 
on the effective Lagrangian traditionally associated with
the standard non-linear realization of supersymmetry \cite{nl}, 
which was believed at that time to be unique at leading order.
As was recently emphasized \cite{bfzeegg}, however, such an
approach is not equivalent to ours, and the two lead to different 
results, both consistent with supersymmetry. For a model-independent 
study, we would need the general form of the low-energy effective 
interactions, allowed by the non-linearly realized supersymmetry, 
that may contribute to the relevant amplitudes at leading order. 
This was not known until recently, when a definite theoretical 
prescription for such an investigation became available 
\cite{clnew} (the same paper also confirmed the results of 
\cite{bfzeegg}, and found the explicit general form of the 
coupling of two on-shell goldstinos to a single photon). 
We are postponing a general phenomenological analysis to a 
forthcoming paper. Because of the strong and universal power-law 
behaviour of the cross-sections, always proportional to $s^3/
F^4$, we expect our results to be rather stable with respect 
to variations of the parameters characterizing the most general 
non-linear realization. However, should a signal show up at 
the Tevatron or the LHC, having the general expression of the 
cross-section would be very important, since a detailed analysis 
of the spectrum and of the angular distributions of photons, jets
and missing energy would offer the unique opportunity of 
distinguishing among possible fundamental theories.
 
\section{Partonic cross-sections}
We now compute the cross-sections for the processes
of eqs.~(\ref{qqgam})--(\ref{qbarglu}), and comment 
on the cross-section for the process of eq.~(\ref{gluglu}).

For the process $q \ov{q} \to \grav \grav \gamma$, the 
calculation is a trivial extension of the one performed 
in \cite{bfz4} for $e^+ e^- \to \grav \grav \gamma$.
We give here, for future use, a compact expression for 
the matrix element squared, summed over the helicities
of the initial and final states. The calculation has 
been performed both by standard trace techniques, with
the help of the program Tracer \cite{tracer}, and by  
helicity-amplitude techniques~\cite{helamp}. We denote
by $(p_1,p_2,q_1,q_2,k)$ the four-momenta of the incoming
quark and antiquark and of the outgoing gravitinos and photon,
respectively. For any given quark flavour, the result is 
\be
\label{summsq}
\begin{array}{c}
\langle \vert M(q \bar q \to \grav \grav \gamma) \vert^2 \rangle
=
\displaystyle{
\frac{Q_q^2 \alpha}{N}
\frac{32 \pi}{F^4 (k\cdot p_1)
(k\cdot p_2)(p_1\cdot p_2)}}
\\ \phantom{bla} \\
\times \left[
A(p_1,p_2,q_1,q_2,k) + A(p_2,p_1,q_1,q_2,k) 
+ \left( q_1 \leftrightarrow q_2 \right) \right] \, ,
\end{array}
\ee
where the symbol $\langle \ldots \rangle$ denotes the average (sum)
over colours and helicities of the initial (final) state, $Q_q$ is 
the electric charge of the quark under consideration, $\alpha \equiv 
e^2 / (4 \pi)$ is the electromagnetic fine-structure constant, $N=3$
is the number of colours and
\bea
& & 
A(p_1,p_2,q_1,q_2,k) = 
\nn \\ & & 
(k p_1)(k p_2)^2(k q_1)(p_1 p_2)(p_1 q_2) 
- (k p_1)(k p_2)(k q_1)^2(p_1 p_2)(p_1 q_2) 
\nn \\ & & 
- (k p_1)(k p_2)^2(k q_1)(p_1 q_1)(p_1 q_2)
+ (k p_1)(k p_2)(k q_1)^2(p_1 q_1)(p_1 q_2) 
\nn \\ & & 
-  2(k p_1)(k p_2)(k q_1)(p_1 p_2)(p_1 q_2)^2 
+ (k p_1)(k q_1)^2(p_1 p_2)(p_1 q_2)^2 
\nn \\ & & 
+ (k p_1)(k p_2)(k q_1)(p_1 q_1)(p_1 q_2)^2 
+ (k p_1)^2(k p_2)(k q_1)(p_1 q_2)(p_2 q_1)
\nn \\ & & 
- (k p_2)(k q_2)(p_1 p_2)(p_1 q_1)(p_1 q_2)(p_2 q_1) 
+ (k p_2)(k q_2)(p_1 q_1)^2(p_1 q_2)(p_2 q_1) 
\nn \\ & & 
- (k p_1)^2(k q_1)(p_1 q_2)^2(p_2 q_1) 
+ (k p_2)(k q_1)(p_1 p_2)(p_1 q_2)^2(p_2 q_1) 
\nn \\ & & 
- (k p_2)(k q_1)(p_1 q_1)(p_1 q_2)^2(p_2 q_1) 
- (k p_1)(k q_2)(p_1 q_1)(p_1 q_2)(p_2 q_1)^2 
\nn \\ & & 
+ (k p_2)(k q_2)(p_1 p_2)(p_2 q_1)^3          
- (k q_2)(p_1 p_2)(p_1 q_2)(p_2 q_1)^3 
\nn \\ & & 
+ (k p_1)(k q_1)(p_1 q_1)(p_1 q_2)(p_2 q_1)(p_2 q_2) 
+ (k q_1)(p_1 p_2)(p_1 q_2)(p_2 q_1)^2(p_2 q_2) 
\nn \\ & & 
+ (k p_1)(k p_2)(p_1 p_2)(p_1 q_2)(p_2 q_1)(q_1 q_2) 
- (k p_1)(k q_1)(p_1 p_2)(p_1 q_2)(p_2 q_1)(q_1 q_2) 
\nn \\ & & 
- (k p_1)(k p_2)(p_1 q_1)(p_1 q_2)(p_2 q_1)(q_1 q_2) 
+ (k p_1)^2(p_1 q_2)(p_2 q_1)^2(q_1 q_2) 
\nn \\ & & 
- (k p_2)(p_1 p_2)(p_1 q_2)(p_2 q_1)^2(q_1 q_2) 
+ (p_1 p_2)(p_1 q_2)^2(p_2 q_1)^2(q_1 q_2) \, .
\label{matrix}
\eea
After integrating over part of the phase space, we obtain:
\be
{d^{\, 2} \sigma \over d \xg d \cos \theta_{\gamma}}
=
{Q_q^2 \alpha  \over N}
{s^3 \over 160 \pi^2 F^4}
\cdot f ( \xg, \ctg) \, ,
\label{dsdxdct}
\ee
where $(\sqrt{s},x_{\gamma},\theta_{\gamma})$ are the total energy, 
the fraction of beam energy carried by the photon, and the photon 
scattering angle with respect to the quark, in the centre-of-mass 
frame of the incoming partons. The function $f$ is given by 
\cite{bfz4}
\be              
f ( \xg, \ctg)  
= 
(1 - \xg)^2 \left[ {(1 - \xg)  (2 - 2 \xg + \xg^2) \over \xg \sqtg}  
+ {\xg (-6 + 6 \xg + \xg^2) \over 16} - {\xg^3 \sqtg \over 32} \right]  
\, .
\label{fxct}
\ee

The extension to the process $q \ov{q} \to \grav \grav g$ is 
straightforward, since the only difference is the replacement of 
the photon with the gluon. By replacing $Q_q^2 \, \alpha$ with 
$C_F \, \alpha_S$ in eqs.~(\ref{summsq}) and (\ref{dsdxdct}),
where $C_F=(N^2-1)/2N=4/3$, we obtain:
\be                      
{d^{\, 2} \sigma \over d x_g d \cos \theta_g}
=
{C_F \alpha_S  \over N}
{s^3 \over 160 \pi^2 F^4}
\cdot f ( x_g, \cos \theta_g) \, .
\label{dsdxgdctg}
\ee

For the crossed channels, $q g \to \grav \grav q$ ($\ov{q} g \to 
\grav \grav \ov{q}$), analogous results can easily be obtained by
taking eqs.~(\ref{summsq}) and (\ref{matrix}), by replacing $Q_q^2 
\alpha/N$ with $C_F \alpha_S/(N^2-1)$, by performing the substitution
$k \leftrightarrow -p_2$ ($k \leftrightarrow -p_1$), and by changing 
the overall sign. After integrating over part of the phase space, we 
easily obtain the one-parton inclusive distributions in terms of the 
partonic centre-of-mass beam-energy fraction $x_q (x_{\ov{q}})$ and 
scattering angle $\theta_q (\theta_{\ov{q}})$. For the $q g \to \grav 
\grav q$ channel:
\be                                                   
\frac{d\sigma}{d\xq\, d\ctq} = 
\frac{C_F \alpha_S }{N^2-1} \;
\frac{s^3}{160  \pi^2  F^4} 
\; g(\xq,\ctq) \, ,         
\ee                                                           
where $\theta_{q}$ is the angle between the direction of the incoming and
outgoing quarks, and
\bea
\label{gfunct}
g(\xq,\ctq) & = & (1 - \xq)^2 \left[
\displaystyle{1 + \xq^2 \over 16 \sin^2(\theta_q/2)}
+                       
\displaystyle{( 1 - \xq)(1-2 \xq+ 2 \xq^2) \over 4 \cos^2 (\theta_q/2)} 
\right. \nonumber \\
& + & 
\left. \displaystyle{\xq (12 - 27 \xq + 12 \xq^2) \over 32} 
+ \displaystyle{\xq^2 (3 - 4 \xq) \ctq \over 32} \right]
\, .
\eea                                                     
Because of the obvious symmetry properties of eq.~(\ref{summsq}),
an identical expression is obtained for the $\ov{q} g \to \grav 
\grav \ov{q}$ channel.

It is easy to verify that all the above results satisfy                     
the required factorization properties in the case of 
collinear emission:
\bea
   \frac{d\sigma}{d\xgl\, d\ctgl}(q \bar q \to \grav \grav g) 
    & {\buildrel{g \parallel q}\over{\to}}        &
    \sigma(q \bar q \to \grav\grav; \;(1-\xgl)s) \; \frac{\alpha_S}{\pi}
   \frac{1}{\sqtgl} \; P_{gq}(\xgl)  \; ,\\         
   \frac{d\sigma}{d\xq\, d\ctq}(q g \to \grav \grav q^\prime) 
    & {\buildrel{q^\prime \parallel q}\over{\to}}        &
    \sigma(g g \to \grav\grav; \;(1-\xq)s) \; \frac{\alpha_S}{\pi}
   \frac{1}{4\sin^2(\theta_q/2)} \; P_{qq}(\xq)  \; , \\         
   \frac{d\sigma}{d\xq\, d\ctq}(q g \to \grav \grav q^\prime) 
    & {\buildrel{q^\prime \parallel g}\over{\to}}        &
    \sigma(q \bar q \to \grav\grav; \;(1-\xq)s) \; \frac{\alpha_S}{\pi}
   \frac{1}{4\cos^2(\theta_q/2)} \; P_{qg}(\xq)  \; , 
\eea
where $P_{ij}$ are the standard Altarelli-Parisi splitting kernels, and
\bea
\sigma(q \bar q \to \grav\grav; \; s) &=& \frac{1}{N} \;
\frac{s^3}{160 \pi F^4} \; ,
 \\
\sigma(g g \to \grav\grav; \; s) &=& \frac{1}{N^2-1}\;  
\frac{s^3}{640 \pi F^4} \; .
\eea                                                   
We did not perform the complete calculation of the $g g \to \grav
\grav g$ process, since the contribution of the $gg$ channel is 
significantly suppressed at the highest values of $s$, where the 
signal is more likely to emerge over the QCD backgrounds. To verify 
this statement, we shall use in the next section an approximate 
estimate of the $gg$ process, based on the collinear approximation:
\be 
\frac{d\sigma}{d\xgl\, d\ctgl}(g g \to \grav \grav g) \sim
\sigma(g g \to \grav\grav; \;(1-\xgl)s) \; \frac{\alpha_S}{\pi} 
\; \frac{1}{\sqtgl} \; P_{gg}(\xgl)  \; .                             
\ee                 
Our results will confirm that indeed the contribution of this 
channel is strongly suppressed in the kinematical regions of interest. 

\section{The jet-plus-missing-energy signal}
In this section, we evaluate the rates for the production of a single 
jet of large transverse energy ($E_T$) plus a gravitino pair in hadronic 
collisions. We shall consider the cases of the currently available 
Tevatron data (\tevI, $p \bar p$ collisions at $\sqrt{S}=1.8$~TeV, 
with an integrated luminosity of approximately $100 \pb^{-1}$), of 
the upcoming upgraded Tevatron run (\tevII, $p\bar p$ collisions at 
$\sqrt{S}=2$~TeV, with an integrated luminosity of approximately 
$2 \fb^{-1}$), and of the LHC ($p p$ collisions at $\sqrt{S}=14 \tev$, 
with an integrated luminosity of approximately $10 \fb^{-1}$). The 
dominant irreducible background to the signal comes from the associated 
production of a $Z$ vector boson and a jet, with the $Z$ decaying 
to a neutrino pair. Given the large values of $E_T$ we will
be considering (in excess of $100 \gev$), no significant detector 
backgrounds are expected. 
Notice that the standard experimental  searches of supersymmetry using
multi-jet-plus-$\met$  final states require the presence of at least three
jets. No experimental study has been published, to date, of the
single-jet-plus-$\met$ signal discussed here. As a result we have to rely on
our own background estimate, since nothing is available from the experimental
studies. We shall therefore extract the region of sensitivity  to the value of
the scale $\sqrt{F}$ from the comparison of the expected signal  and the
calculated background rates.

In our numerical analysis, we shall use the parton-density 
parametrization MRSR2~\cite{Martin96}, with a 
renormalization/factorization scale $\mu=E_T$. Since the cross-sections for both
signal and background processes scale linearly with $\alpha_S$, 
their ratio is rather insensitive to this choice. In particular, 
since the signal rate is proportional to $F^{-4}$, the limits 
we shall set on $\sqrt{F}$ are not significantly affected by this choice, as
we have explicitly verified by varying the scale $\mu$ in the 
range $E_T/4 < \mu < 4 E_T$, independently for signal and background.

\begin{figure}
\begin{center}
\epsfig{file=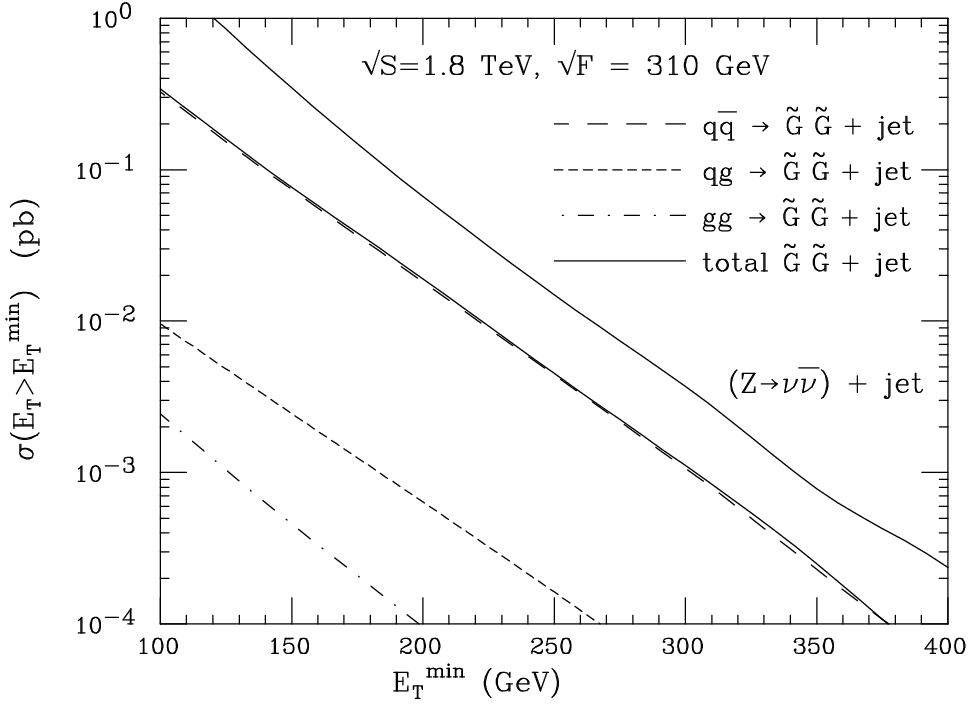,width=0.8\textwidth,clip=}
\end{center}        
\vspace*{-1cm}
\ccaption{}{\label{fig:tev1sig} 
Production rates for $p \bar p \to 
\grav\grav+{\rm jet}$ and $p \bar p \to {\rm jet}+ \nu \bar{\nu}$ at 
$\sqrt{S}=1.8$~TeV.}                    
\end{figure}
\begin{figure}
\begin{center}
\epsfig{file=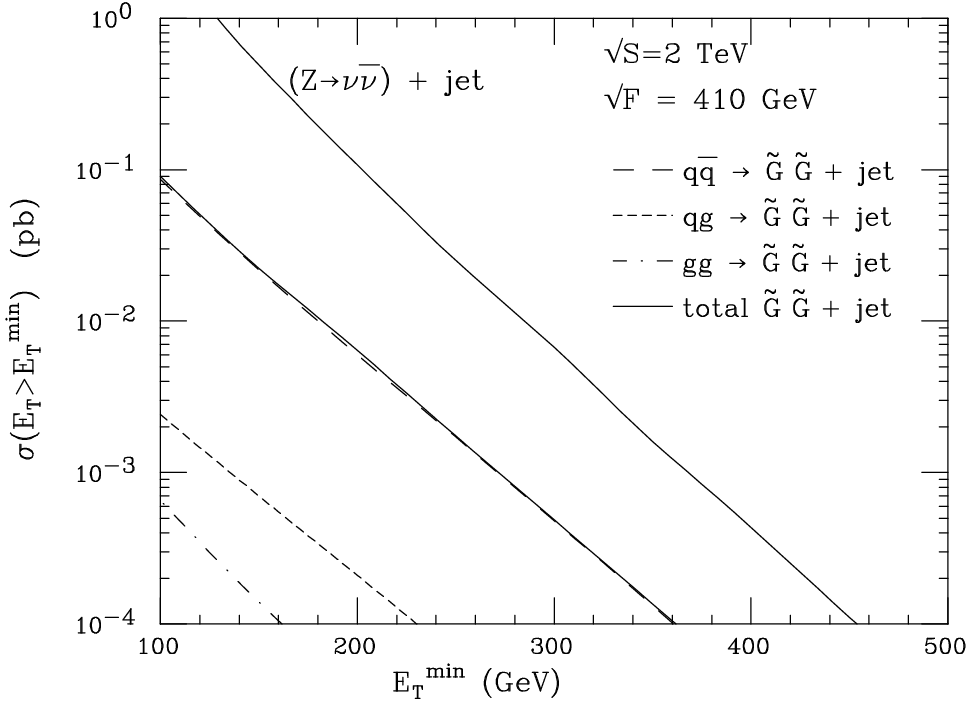,width=0.8\textwidth,clip=}
\end{center}
\vspace*{-1cm}
\ccaption{}{\label{fig:tev2sig} 
Production rates for $p \bar p \to
\grav\grav+{\rm jet}$ and $p \bar p \to {\rm jet} + \nu \bar{\nu}$ at 
$\sqrt{S}=2$~TeV.}                        
\end{figure}
\begin{figure}
\begin{center}
\epsfig{file=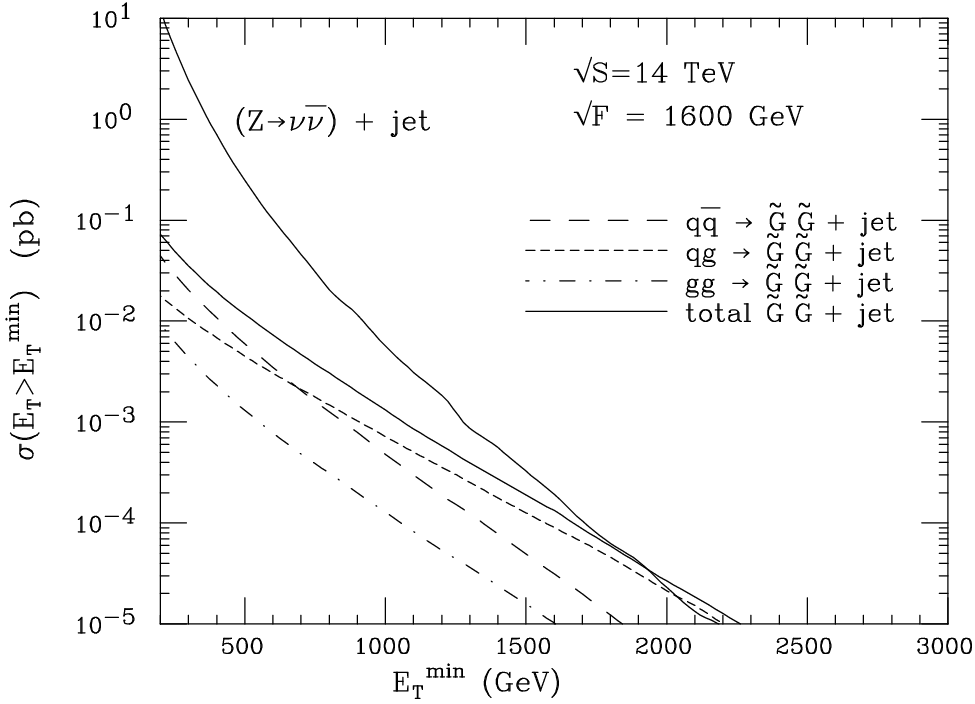,width=0.8\textwidth,clip=}
\end{center}
\vspace*{-1cm}
\ccaption{}{\label{fig:lhcsig} 
Production rates for $p  p \to 
\grav\grav+{\rm jet}$ and $p  p \to {\rm jet} + \nu \bar{\nu}$ at 
$\sqrt{S}=14$~TeV.}                   
\end{figure}
Figures~\ref{fig:tev1sig}--\ref{fig:lhcsig} show the contribution 
to the signal rate of the three possible initial states $q\bar q$, 
$qg+gq$ and $gg$. The distributions shown correspond to the jet 
rate integrated over transverse energies larger than a given 
threshold $\etmin$, and within a jet-rapidity range $\vert \eta
\vert < 2.5$. For the three figures we selected the values of 
$\sqrt{F}=310, \, 410$, and 1600~GeV, respectively. This particular 
choice will be justified later on. The relative contributions of the
three channels and the shapes of their distributions are not 
affected by the value of $\sqrt{F}$. For comparison, we also show the
background rate for the process $p {\buildrel{(-)}\over{p}} \to Z$~jet$\to
\nu\bar{\nu}$~jet. This was calculated at leading order in QCD. The capability
of LO QCD to describe this process is proved by the recent study of the
$Z+$~multijet final states published by CDF~\cite{cdf}, with the $Z$ decaying
to charged lepton pairs. In this paper the ratio Data/Theory for the 
$Z$-plus-1-jet final state was found to be $1.29 \pm 0.17$, using $\mu=\et$.
For consistency with this result, we rescaled our LO background prediction by
a factor 1.29. To be conservative, however, we did not rescale the signal
rates. 
                                                                               
We emphasize the following features of these distributions. First of all,
notice that the shape of the background is very similar to that of the
signal. This suggests that the best limit on $\sqrt{F}$ will be extracted 
from a comparison of signal and background in the widest possible
region of $\et$ compatible with other experimental requirements.
Secondly, notice that the contribution of the $gg$ initial state 
is always negligible in the interesting regions of $\et$, due to the 
smallness of the gluon density compared with the valence quark density.
This is true regardless of the collider energy and justifies our use 
of an approximate expression for the $gg\to \grav\grav g$ matrix 
elements. Finally, notice that while at the Tevatron the signal is 
dominated by the $q \bar q$ initial state, the dominant contribution 
at the LHC comes from the $qg+gq$ channel. This is because the antiquark 
density in $pp$ collisions is much smaller than the gluon density.                                                                       

\begin{figure}
\begin{center}
\epsfig{file=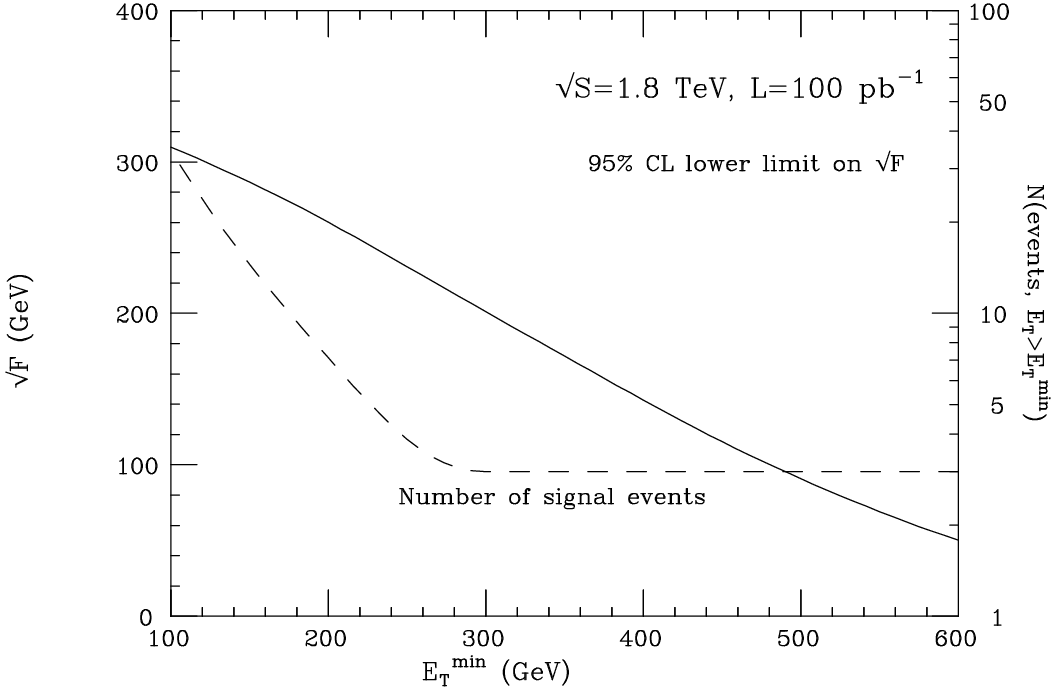,width=0.8\textwidth,clip=}
\end{center}
\vspace*{-1cm}
\ccaption{}{\label{fig:tev1cl} 
Solid line (scale on the left axis):
estimated 95\%~CL exclusion limits on $\sqrt{F}$ from the absence of a deviation
from the SM jet+$\met$ rate in events with jet-$\et>\etmin$ at \tevI.
Dashed line (scale on the right axis): number of ${\rm jet}+
\grav\grav$ events with jet-$\et>\etmin$ corresponding to the 95\%~CL value of
$\sqrt{F}$.}
\end{figure}
\begin{figure}
\begin{center}
\epsfig{file=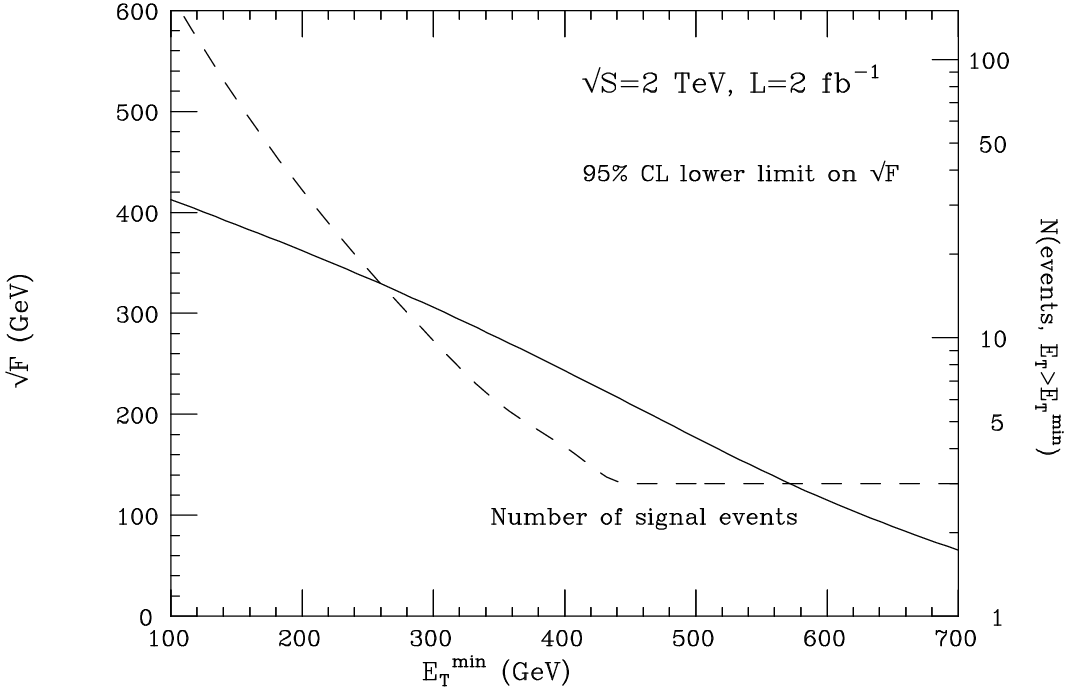,width=0.8\textwidth,clip=}
\end{center}
\vspace*{-1cm}
\ccaption{}{\label{fig:tev2cl}   
Same as fig.~\ref{fig:tev1cl}, for \tevII.}
\end{figure}
\begin{figure}
\begin{center}
\epsfig{file=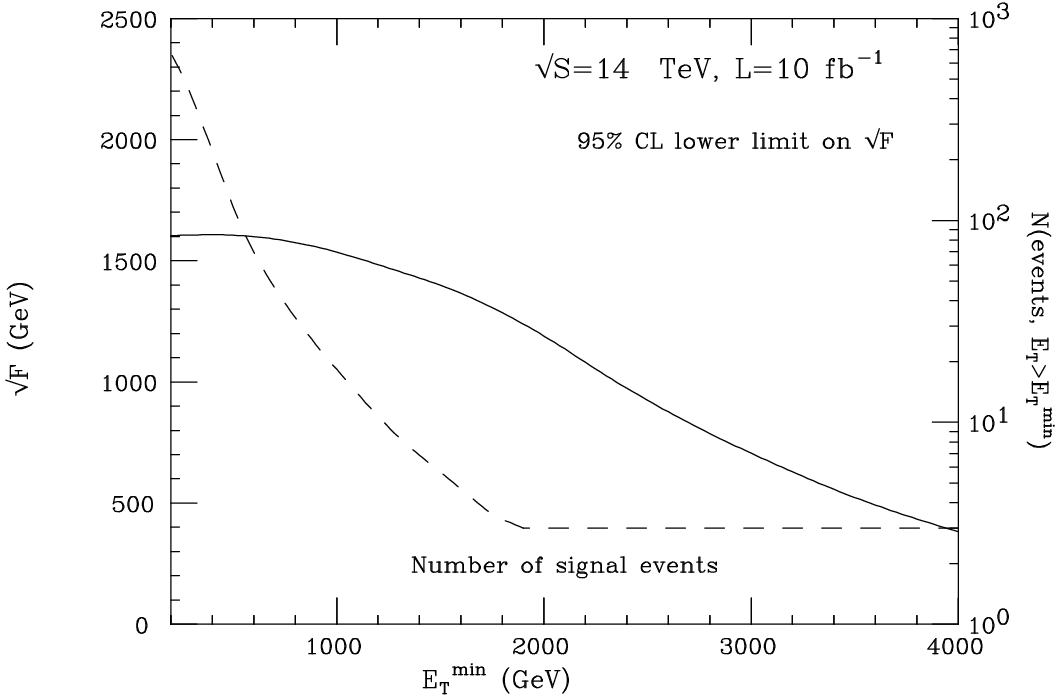,width=0.8\textwidth,clip=}
\end{center}
\vspace*{-1cm}
\ccaption{}{\label{fig:lhccl} 
Same as fig.~\ref{fig:tev1cl}, for the LHC.}
\end{figure}

To extract the region of sensitivity to the scale $\sqrt{F}$, we 
count the number of background events expected above a given 
threshold $\etmin$ and evaluate the value of $\sqrt{F}$ that can 
be excluded at the $95\%$ confidence level (CL). For values of 
$\etmin$ such that no more background events are expected, this
corresponds to a limit of 3 signal events. The $95\%$~CL lower 
limits on $\sqrt{F}$ for the three cases considered are plotted 
as functions of $\etmin$ in figs.~\ref{fig:tev1cl}--\ref{fig:lhccl}.
In the same figures we also plotted (as dashed lines) the number of 
signal events corresponding to the selected value of $\sqrt{F}$. The 
best limits on $\sqrt{F}$ that can be extracted with the chosen 
integrated luminosities are given by 310, 410 and 1600 GeV for \tevI, 
\tevII~and the LHC, respectively. The corresponding limits on $m_{3/2}$ 
are $2.3 \times 10^{-5} \ev$, $4.0 \times 10^{-5} \ev$ and $6.2 \times 
10^{-4} \ev$. These limits are obtained for $\etmin = 100 \gev$ in the
case of \tevI~and \tevII, and for $\etmin = 200 \gev$ in the case of
the LHC. Lower values of $\etmin$ could lead to more stringent limits,
but in this case other backgrounds could become important, and a 
detailed experimental analysis would be required. We should also mention
that our procedure relies on the detailed prediction of the background 
rates. Notice nevertheless that in the regions of $\et$ where a large 
background is expected, its absolute normalization can be determined 
with high accuracy by measuring the $(Z\to \ell^+\ell^-)+{\rm jet}$ rates.
In these regions of large rates, furthermore, one could use not just the
total rate above a given threshold, but the $E_T$-distribution as well. 

\section{The photon-plus-missing-energy signal}
\begin{figure}
\begin{center}
\epsfig{file=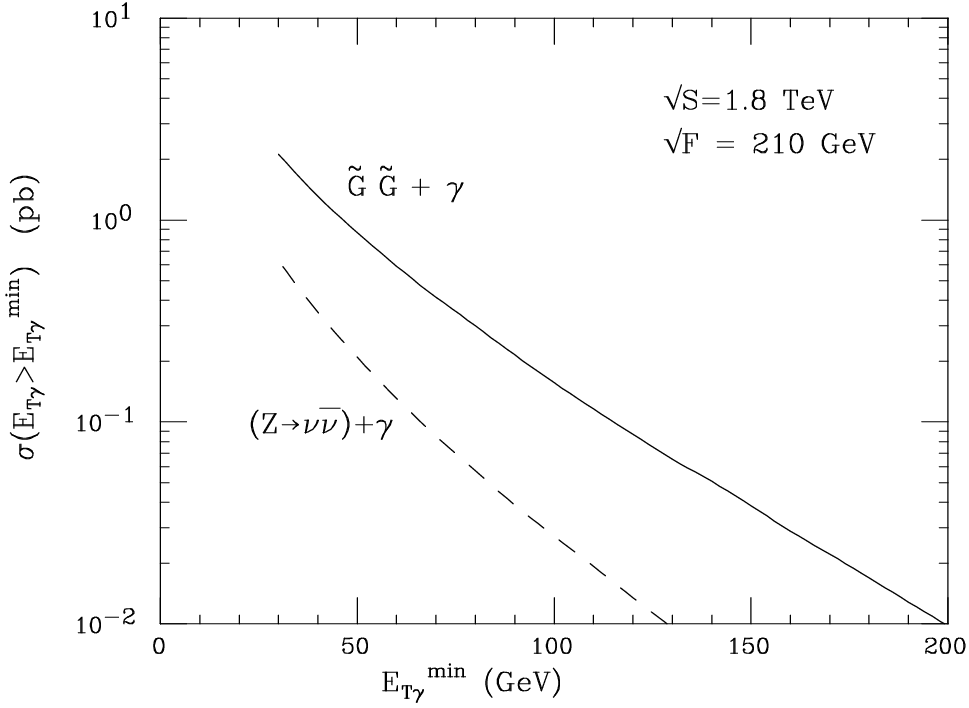,width=0.8\textwidth,clip=}
\end{center}
\vspace*{-1cm}
\ccaption{}{\label{fig:tev1gsig} 
Production rate for $p \bar p \to
\grav\grav+\gamma$ at $\sqrt{S}=1.8$~TeV.}
\end{figure}
\begin{figure}
\begin{center}
\epsfig{file=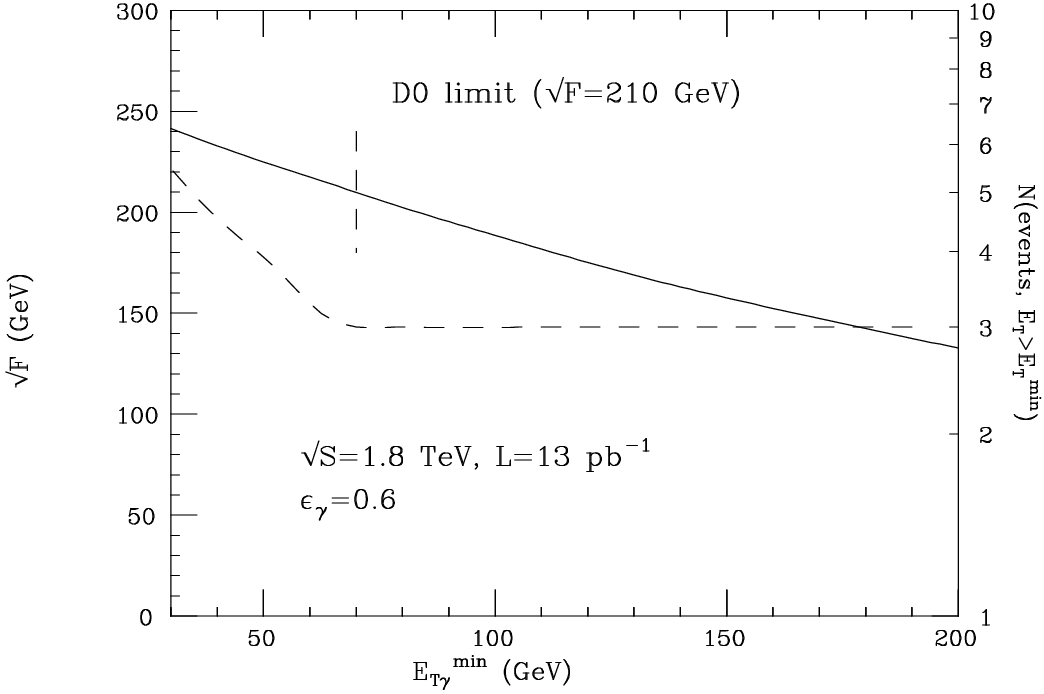,width=0.8\textwidth,clip=}
\end{center}
\vspace*{-1cm}
\ccaption{}{\label{fig:tev1gcl} 
95\%~CL exclusion limits on $\sqrt{F}$ from the absence of 
$\gamma+\met$ events at D0~\cite{d0}.}
\end{figure}
While the production rate of events with a photon plus the 
gravitino pair is smaller than that with a jet, the relevant 
source of background (associated production of $Z \gamma$) is 
also smaller. A recent analysis from D0~\cite{d0}, for example, 
did not find any $\gamma+\met$ event with $E_{T\gamma}>70 \gev$
in approximately 13~pb$^{-1}$ of data. It is therefore interesting 
to examine the constraints on $\sqrt{F}$ set by searches in this 
channel. Figure~\ref{fig:tev1gsig} shows the $\gamma+\met$ signal 
rate, as a function of the photon $\et$ threshold. We chose for 
this plot $\sqrt{F}=210$~GeV. In the case of photon final states, 
only the $q \bar q$  annihilation channel contributes. The results 
we find agree with the naive estimate:
\be
\frac{\sigma(\grav \grav \; \gamma)}{\sigma(\grav \grav \; 
{\rm jet})} \sim \frac{Q_u^2 \alpha}{C_F \alpha_S} \sim 3 \% \; .
\ee                                                             
The 95\%~CL exclusion limit on $\sqrt{F}$ as a function of $\etmin$, 
for the 13~pb$^{-1}$ of data corresponding to the available D0 
analysis, is given in fig.~\ref{fig:tev1gcl}. In addition to 
imposing a geometrical acceptance cut of $\vert \eta_{\gamma} 
\vert<2.5$, we reduced both signal and background rates by the 
photon identification efficiency $\epsilon_{\gamma}=0.6$~\cite{d0}. 
For $\etmin=70$~GeV, the threshold value above which D0 reports 
no events found, we obtain $\sqrt{F}>210$~GeV. Slightly better 
constraints could be obtained by using lower thresholds. We look 
forward to the completion of the analyses of the full 100~pb$^{-1}$ 
current data sample, for which we estimate a 95\%~CL limit of 250~GeV 
(265~GeV) for $\etmin=70$~GeV (50~GeV). These numbers are only a factor 
of 15\% worse than those obtained with the jet+$\met$ analyses. 
Further improvements may arise from a combination of the D0 and CDF 
statistics.
                                               
\begin{figure}
\begin{center}
\epsfig{file=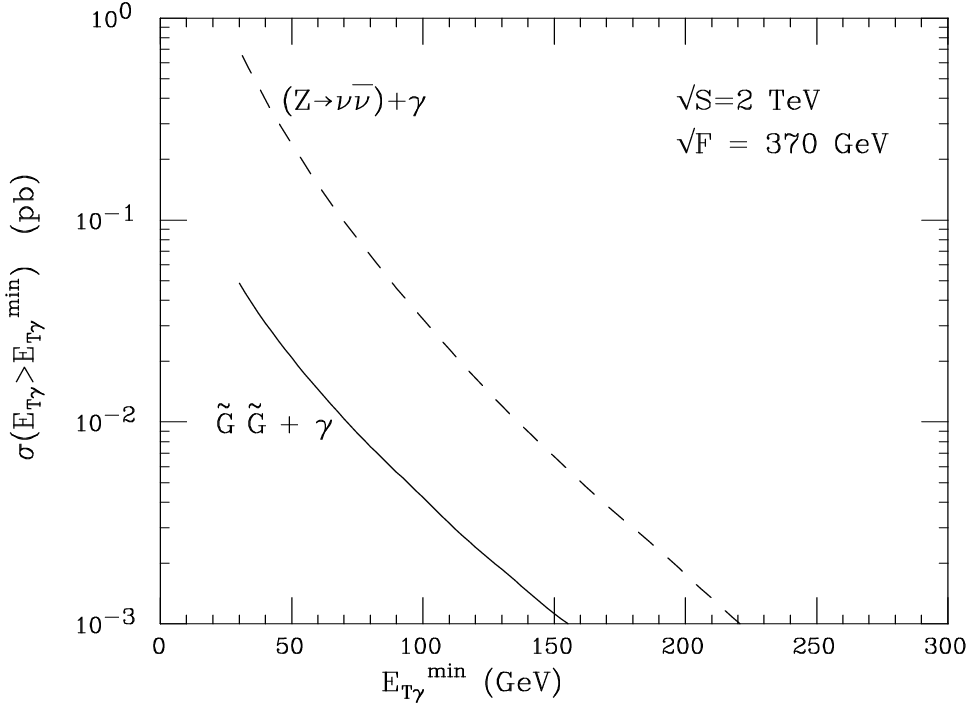,width=0.8\textwidth,clip=}
\end{center}
\vspace*{-1cm}
\ccaption{}{\label{fig:tev2gsig} 
Production rate for $p \bar p \to 
\grav\grav + \gamma$ at $\sqrt{S}=2$~TeV.}
\end{figure}
\begin{figure}
\begin{center}
\epsfig{file=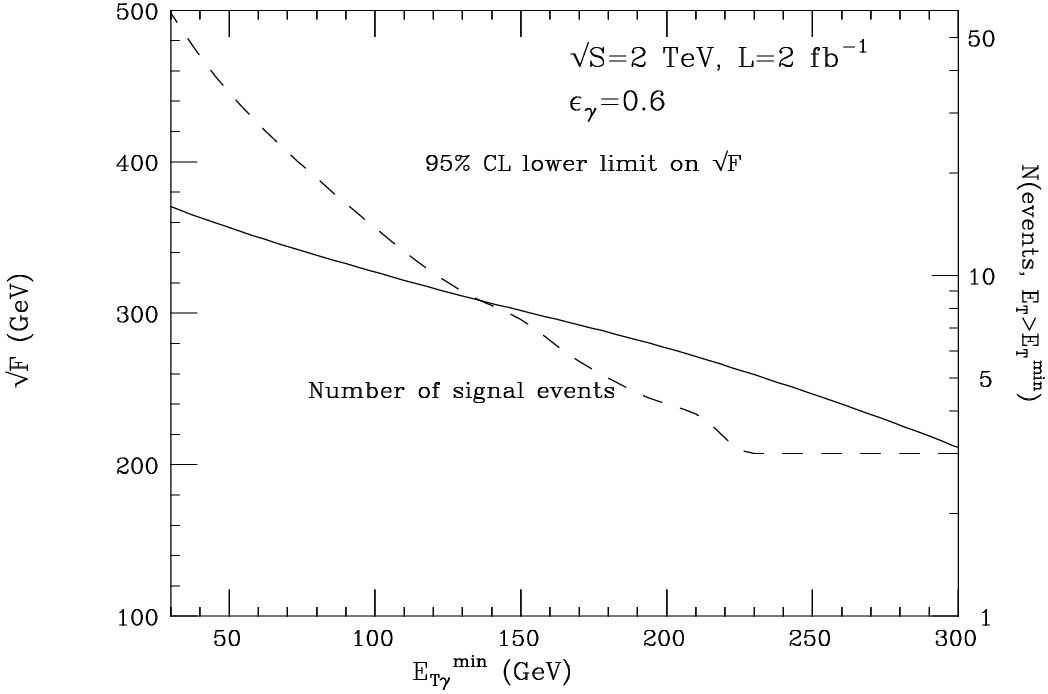,width=0.8\textwidth,clip=}
\end{center}
\vspace*{-1cm}
\ccaption{}{\label{fig:tev2gcl} 
Solid line (scale on the left axis):
estimated 95\%~CL exclusion limits on $\sqrt{F}$ from the absence 
of a deviation from the SM $\gamma+\met$ rate in events with 
$\gamma$-$\et>\etmin$ at \tevII.
Dashed line (scale on the right axis): number of $
\grav\grav+\gamma$ events with $\gamma$-$\et>\etmin$ corresponding to the 
95\%~CL value of $\sqrt{F}$.}
\end{figure}
\begin{figure}
\begin{center}
\epsfig{file=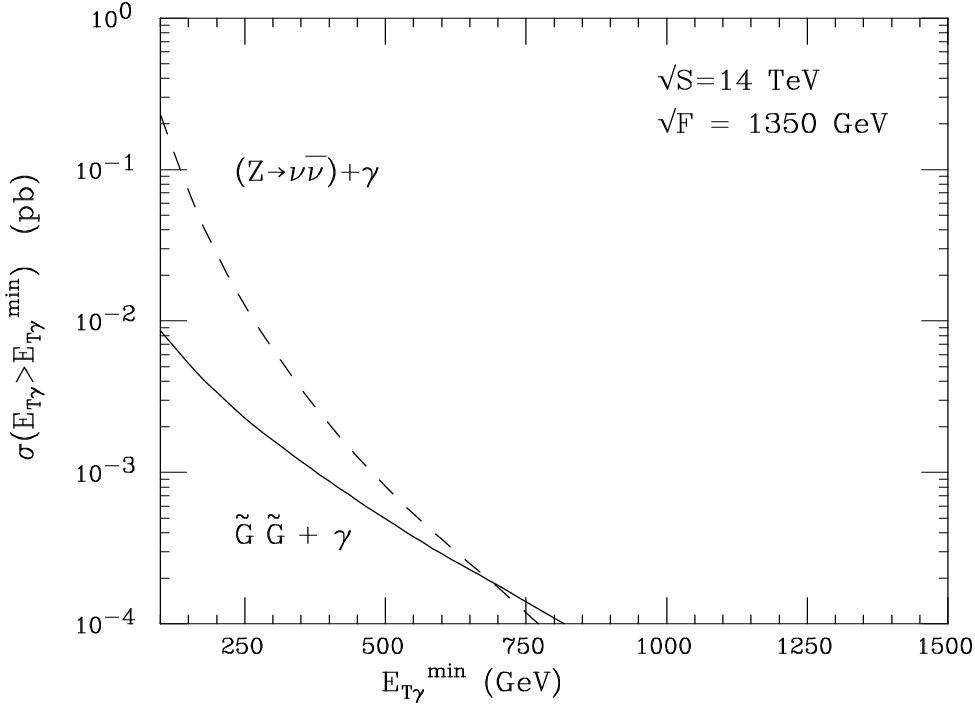,width=0.8\textwidth,clip=}
\end{center}
\vspace*{-1cm}
\ccaption{}{\label{fig:lhcgsig} 
Production rate for $p p \to \grav\grav+\gamma$ at $\sqrt{S}=14$~TeV.}
\end{figure}
\begin{figure}
\begin{center}
\epsfig{file=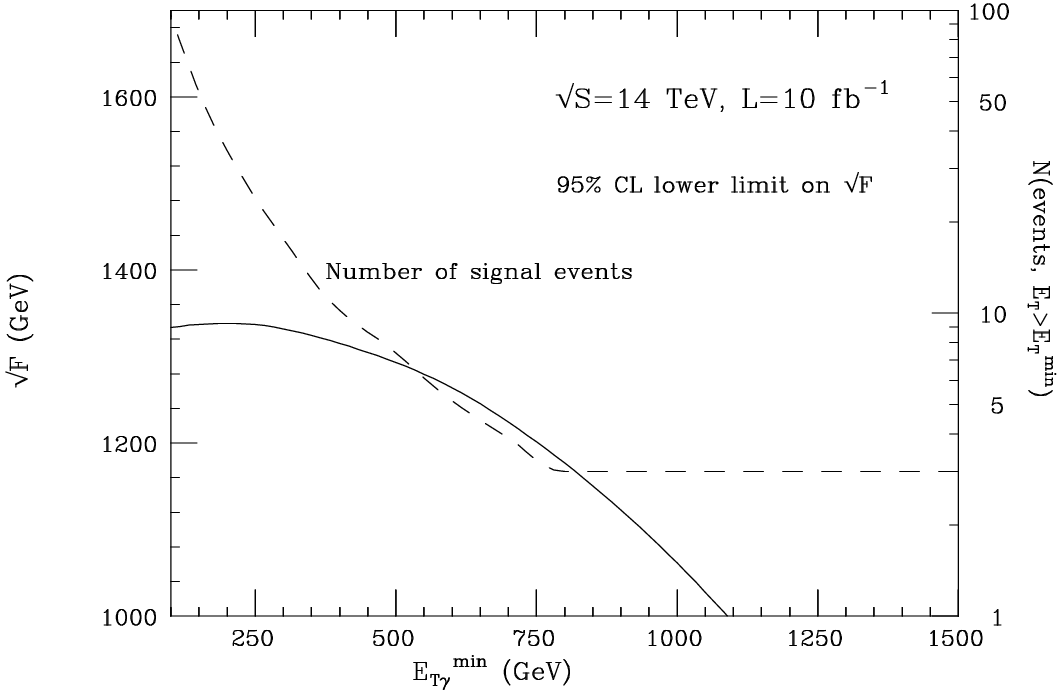,width=0.8\textwidth,clip=}
\end{center}
\vspace*{-1cm}
\ccaption{}{\label{fig:lhcgcl} 
Solid line (scale on the left axis): estimated 95\%~CL exclusion limits 
on $\sqrt{F}$ from the absence of a deviation
from the SM $\gamma+\met$ rate in events with 
$\gamma$-$\et>\etmin$ at the LHC.
Dashed line (scale on the right axis): number of $
\grav\grav+\gamma$ events with $\gamma$-$\et>\etmin$ corresponding to the 
95\%~CL value of $\sqrt{F}$.}
\end{figure}

The sensitivity achievable in the $\gamma+\met$ channel at \tevII\ is 
displayed in figs.~\ref{fig:tev2gsig} and \ref{fig:tev2gcl}, which show 
the production rates and 95\%~CL limits after accumulation of 2~fb$^{-1}$ 
of data at 2~TeV. Similar plots for the LHC are shown in 
figs.~\ref{fig:lhcgsig} and \ref{fig:lhcgcl}. In this case we set the 
photon-identification efficiency equal to 1, in the absence of a 
determination based on real data. While the Tevatron sensitivity in the 
photon channel is almost comparable to that in the jet channel, a clear 
advantage of the jet signal over the photon one is seen at the LHC. This 
is due to the smaller luminosity of the $q\bar q$ initial state relative 
to the $qg$ initial state at the LHC.

\begin{figure}
\begin{center}
\epsfig{file=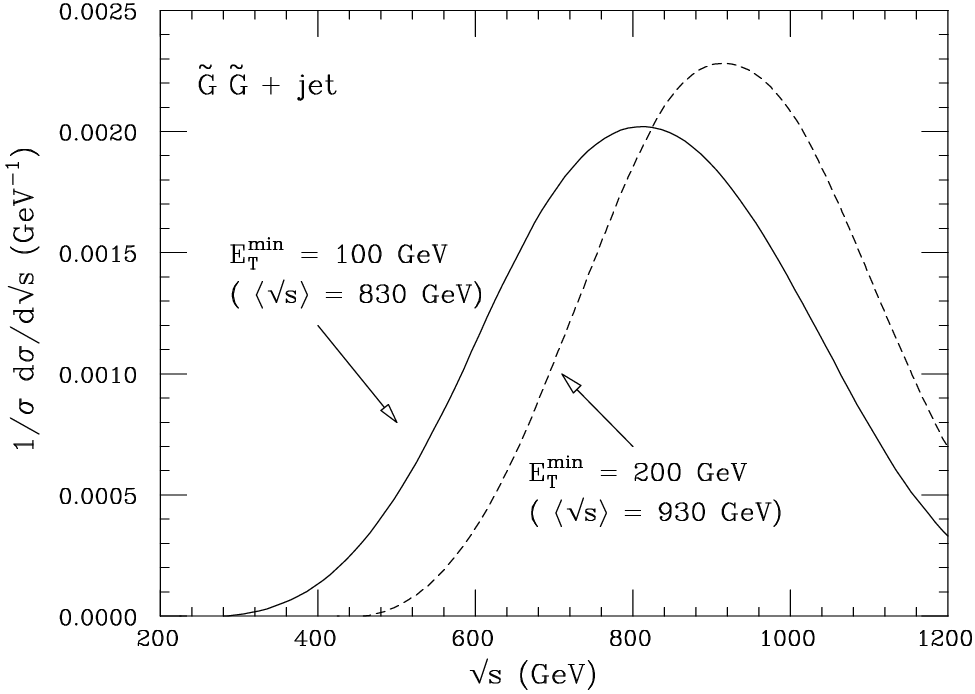,width=0.8\textwidth,clip=}
\end{center} 
\vspace*{-1cm}
\ccaption{}{\label{fig:shat} 
$\sqrt{s}$ distribution for events with $\et>100$~GeV (solid line) and
$\et>200$~GeV (dashed line) at the Tevatron.}        
\end{figure}

\section{Discussion and conclusions}
We presented in this paper a phenomenological analysis of high-energy 
collider constraints on the scale of supersymmetry breaking  $\sqrt{F}$, 
in one-to-one correspondence with the gravitino mass $m_{3/2}$, in 
models where the only light supersymmetric particle is the gravitino. 
A typical signature for these models
is the annihilation of $e^+e^-$ or $q \ov{q}$ pairs into gravitino pairs. 
To tag the events, additional emission of either a jet or a photon is 
necessary. The study 
of LEP data in ref.~\cite{bfz4} has been extended here to signals from 
the Tevatron collider. The limits from the Tevatron are significantly 
stronger than those obtained from LEP, thanks to the higher energy reach. 
Supersymmetry-breaking scales as large as $\sqrt{F} = 210$~GeV ($m_{3/2}
=1.1 \times 10^{-5}$~eV) are already excluded at 95\%~CL by a D0 search 
for events with photons and missing energy. The analyses of the full 
\tevI\ data sample should increase this limit to $\sqrt{F} \simeq 265 
\gev$ ($m_{3/2} \simeq 1.7 \times 10^{-5} \ev$). 
Comparable and probably better limits should be obtained from the current 
data in the jet plus $\met$ final state, where we foresee sensitivity up 
to $\sqrt{F} \simeq 310 \gev$ ($m_{3/2} \simeq 2.3\times 10^{-5}$~eV). 
The upgraded 
Tevatron, accumulating 2~fb$^{-1}$ at 2~TeV after the year 2000, should 
improve the sensitivity up to $\sqrt{F} \simeq 410 \gev$ ($m_{3/2} \simeq 
4.0 \times 10^{-5}$~eV). The sensitivity will then increase to $\sqrt{F}
\simeq 1.6 \tev$ ($m_{3/2} \simeq 6.2 \times 10^{-4} \ev$) once the 
LHC data become available.  
                                                          
These estimates have been obtained by using an effective Lagrangian 
containing only the degrees of freedom associated to the ordinary 
particles and the $\pm 1/2$ helicity states of the gravitino.
Such a Lagrangian is non-renormalizable and is expected to provide 
a good approximation of the fundamental theory in an energy range  
bounded from above by a critical value $E_c \sim (2 \div 3) \sqrt{F}$. 
Beyond $E_c$, perturbative unitarity gets violated and the non-local
character of the low-energy theory should show up. These considerations 
also bound the mass $M$ of the new particles called to rescue unitarity:
$M \simlt (2 \div 3) \sqrt{F}$. We should therefore check that, for each 
reaction considered, the typical energy $\sqrt{s}$ probed by the partonic 
process remains within the allowed domain. Similarly, to be consistent, 
we should be able to consider superpartner masses sufficiently large 
to suppress direct production or large threshold effects, while 
remaining within a few times the supersymmetry breaking scale 
$\sqrt{F}$~\footnote{We recall that the complete absence of supersymmetric 
particles with masses below the TeV scale is generically disfavoured by 
naturalness considerations, even if moderate exceptions may exist. For 
this reason, the limit on $\sqrt{F}$ obtained from our study of the future 
LHC data essentially saturates the range of applicability of our formalism.
Note also that our approach does not set interesting constraints on 
models with gauge-mediated supersymmetry breaking (for a review
see ref.~\cite{gr}), in which typical superpartner masses are 
much smaller than $\sqrt{F}$, so that one is lead to set $\sqrt{F}
\gsim 10-100$~TeV.}.                            
                                                             
These constraints turn out to hold for the cases considered in this 
paper. As an example, we show in fig.~\ref{fig:shat} the distribution 
of the variable $\sqrt{s}$ for jet+$\met$ events with $\etmin=100$ and
200~GeV at the Tevatron. The figure shows that, for the typical 
sensitivity range of $\sqrt{F}\sim 300$~GeV, the average value of 
$\sqrt{s}$ probed by this class of events is indeed of the order 
of $(2 \div 3) \, \sqrt{F}$. We verified that similar results hold at 
the LHC, with average values of $\sqrt{s}$ in the range of 3~TeV.
                              
The study presented in this paper should allow experimentalists
to establish absolute lower bounds on $\sqrt{F}$ and $m_{3/2}$.
Indeed, we do not expect that our bounds dissolve as the threshold 
for superparticle production is approached or crossed. On the contrary,
in that case possibly stronger limits on the gravitino mass could be
put~\cite{dicus} by exploiting the production of the gravitino in 
association with squarks and gluinos, or the production of jets in 
association with the spin--$0$ partners of the goldstino.  No 
conclusive phenomenological study of this type has however been 
carried out as yet, since no experimental analyses are available 
of final states with one jet plus $\met$.
                                                  
The reader may have noticed that, in our analysis, all
weak-interaction effects were neglected. For the signals 
discussed, and in the case of a goldstino neutral under 
the full electroweak gauge group, this is a consistent 
approximation: virtual $W$ and $Z$ exchange cannot give
appreciable contributions. We may ask, however, if the
associated production of a gravitino pair and a weak vector
boson may lead to detectable signals. For the $\grav \grav 
Z$ final state, the Standard Model background ($Z Z$ production
with one of the $Z$ decaying to a neutrino pair) is very small: 
only a few $Z Z$ events have been observed up to now at \tevI.
The signal, however, is also expected to be strongly suppressed
with respect to the photon case, since the gain due to the coupling 
constant enhancement is much less than the loss due to the small 
leptonic branching ratio (hadronic decays suffer from severe 
background problems) and to the effects of the finite $Z$ mass. 
For the $\grav \grav W$ final state the background is large,
being dominated by production of a single off-shell $W$.
In both cases, we expect weaker limits on $\sqrt{F}$ and 
$m_{3/2}$ than the ones discussed in this paper. 
                                       
We may also ask if the process $q \gamma \to \grav \grav
q$, analogous to the ones considered in the present paper,
may give rise to detectable signals at HERA. We studied this 
process in $e p$ collisions at $\sqrt{s} = 300 \gev$. With $100 
\pb^{-1}$, roughly corresponding to the present integrated
luminosity collected by the ZEUS and H1 experiments, we found 
no sensitivity at HERA for values of $\sqrt{F}$ in excess of 
$100 \gev$.

In conclusion, we expect that hadron colliders will lead
the search for a superlight gravitino in the next decade.

%
%
                                      
%
\end{document}